# Enhancement of database access performance by improving data consistency in a non-relational database system (NoSQL)

Adam A. E. Alflahi[1], Mohammed A. Y. Mohammed[2], Abdallah Alsammani[3*]

[1] Department of Computer Science and Mathematics, Bahri University, Khartoum, Sudan.
[2] Department of Mathematics and Statistics, Georgia State University, Atlanta, GA, USA.
[3*] Department of Mathematics, Jacksonville University, Jacksonville, FL, USA.
aalsamm@ju.edu

## Abstract

This study aims to enhance data consistency in NoSQL databases, traditionally designed with BASE properties, as opposed to the strong consistency guaranteed by ACID principles in RDBMS. We introduce a comprehensive four-stage server-side model engineered explicitly for MongoDB. This model covers transaction management, bifurcation of read and write transactions, assessment of transaction readiness, and transaction execution via a specialized locking algorithm. Utilizing the Yahoo Cloud Services Benchmark (YCSB), particularly for update-heavy workloads (A, B, and F), our model exhibited significant improvements. Specifically, the average throughput, read, and update latencies improved to 2864.726 ms, 32806.275 ms, and 51845.629 ms, respectively, from the baseline metrics of 2914.110 ms, 26510.930 ms, and 32457.662 ms. These results demonstrate the efficacy of our proposed model in enhancing consistency not only in document-based NoSQL databases like MongoDB but also in other NoSQL database variants, including key-value, graph, and wide-column stores.

**Keywords:** NoSQL, BASE, RDBMS, ACID, MongoDB, YCSB, benchmark, transaction management, database consistency.

## 1. Introduction

In the digital age, the exponential growth of data and its diversified nature have necessitated the evolution of data storage techniques. While highly structured and reliant on the ACID properties (Atomicity, Consistency, Isolation, Durability) for ensuring data accuracy and reliability, conventional databases often falter when dealing with voluminous and varied data. This inadequacy gave rise to NoSQL databases which prioritize flexibility, scalability, and speed. NoSQL, which stands for "Not Only Structure Query Language," offers an alternative to the rigid structures of traditional relational databases.

NoSQL databases are primarily categorized into four types: Graph Stores, which utilize graph structures such as nodes and edges for data storage; Document Stores that employ BSON (Binary JSON) documents for this purpose; Key-Value Stores, which rely on unique key-value pairs for storing data; and Wide Column Stores that use columns as their primary means of data storage.

The evolution of data storage and retrieval methodologies has witnessed a significant transition from traditional relational databases towards NoSQL systems, primarily influenced by their inherent ability to accommodate the vast and diverse nature of Web 2.0 data [1]. However, these systems deviate from the relational model's foundational consistency measures [2]. While NoSQL databases promise advantages like scalability and flexibility, they introduce challenges in ensuring data consistency, especially in commercial cloud storage [1]. Vogels laid the foundation by discussing the concept of eventual consistency, wherein updates might not be instantaneous but are guaranteed over time [3]. This idea was further extended by researchers who explored flexible update mechanisms for weakly consistent replication, creating a bridge between strict consistency and high availability [4]. Parallel efforts in this arena have seen González-Aparicio et al. designing transaction processes that focus on user applications [5] and Lotfy et al. suggesting a middle layer to support ACID properties traditionally absent in NoSQL systems [6]. Taking a broader perspective, systems like "Megastore" [7] and "CloudTPS" [8] have emerged to offer scalable, consistent storage solutions in distributed environments. The narrative was further enriched by solutions like "Granola," aiming at low overhead in transaction coordination [9], and "ElasTraS," envisioning elasticity in cloud transaction storage [10]. Amidst these developments, a focus on specific tools and their transaction capabilities has flourished, such as exploring MongoDB's scalable transactions [11]. Innovative solutions like policy-based access control for consistency [12] and multilevel transaction models for distributed systems [13] have showcased the depth and breadth of research in reconciling scalability with data consistency.

Efforts to improve consistency in NoSQL databases have been a focus of recent research, given these databases often trade off strong consistency for scalability [14], [15]. Various locking mechanisms have been studied to maintain data consistency in NoSQL systems [16], [17], [18]. Alternative consistency models, like Causal Consistency, have also been explored to balance performance and consistency [16], [19]. Tools like the Yahoo Cloud Serving Benchmark (YCSB) are the standard for evaluating database performance [20]. Consequently, enhancing consistency in scalable NoSQL databases remains an active area of research.

This paper delves deeper into NoSQL databases' challenges, particularly focusing on their inherent trade-offs between scalability and consistency. While NoSQL databases have revolutionized how data is stored and managed, they often operate on the BASE properties, potentially compromising the strong consistency offered by traditional RDBMSs and their ACID properties. In our pursuit to bridge this gap, we introduce a server-side model segmented into four pivotal stages designed to enhance database consistency. This model's efficacy is empirically examined through the YCSB benchmark, targeting workloads that underscore update operations. Our findings illuminate significant improvements in average throughput, read, and update timings post-implementation of our proposed algorithm. Our research demonstrates a promising advancement in achieving enhanced consistency for NoSQL databases. We remain optimistic about the potential applicability of our model across various NoSQL platforms, including key-value, graph, and wide-column stores, thereby pushing the boundaries of what these databases can offer regarding data reliability.

## 2. Methods

Informed by the insights gleaned from our comprehensive literature review, we meticulously crafted a bespoke framework, ensuring a cohesive integration of its components to best address the identified challenges and gaps. This architecture was conceived to seamlessly blend the functionalities of the chosen tools while allowing for scalability and adaptability to potential future modifications.

**Tools Employed:**

**MongoDB Database:** Chosen for its document-centric approach, MongoDB is a leading NoSQL database known for its high performance, scalability, and flexibility. Within our framework, MongoDB plays a pivotal role, tasked with not only securely storing the requisite data but also efficiently handling and executing user transactions. Given the inherent BASE properties of MongoDB, special attention was given to the schema design and transaction management to align with our consistency enhancement objectives.

**Performance Benchmarking with YCSB:** Measuring and validating the effectiveness of any proposed solution is paramount. With this in mind, we incorporated the Yahoo Cloud Services Benchmark (YCSB) tool into our framework. Renowned for its capability to provide real-world, industry-standard performance metrics, YCSB was essential in generating empirical data on our framework's performance. It allowed us to perform a rigorous assessment, offering metrics that vividly showcased the efficiency improvements achieved during various stages of data processing.

**Programming Environment:** Java: To bring our proposed model to fruition, we leveraged Java—a versatile, object-oriented programming language acclaimed for its platform independence and extensive library support. This choice was motivated by Java's ability to deliver robust solutions capable of integrating with diverse tools like MongoDB and YCSB and its proven track record in building scalable, high-performance applications. The development was augmented by using established design patterns and best practices, ensuring that our model was both maintainable and efficient.

## 3. Results

We evaluated the performance impact of our proposed algorithm on multi-document transactions in MongoDB using the YCSB benchmark standards, specifically focusing on workloads A, B, and F. Our primary goal was to assess the read/write consistency and its effect on multi-document performance in MongoDB after integrating our algorithm into the YCSB framework.

To achieve this, we introduced a lock() class within the YCSB benchmark, temporarily locking a MongoDB document accessed by the initial client's transaction. This transaction concludes by invoking the commit() method upon completing the abort() method if not finalized within a set timeframe, resulting in a transaction rollback. Subsequently, the document is unlocked, making it accessible to other clients.

Our experiments were conducted on a dataset consisting of 10,000 operations with 10,000 records. The performance was assessed with and without the proposed algorithm across various client configurations: one, five, ten, and fifteen clients. Each client operated independently, with a single client executing 10,000 operations/ms, five clients managing 50,000 operations/ms, ten clients at 100,000 operations/ms, and fifteen clients performing 150,000 operations/ms.

The benchmark tests followed a two-phase structure: a loading phase, where data was loaded, and a run phase, which executed the data based on the selected YCSB workloads. This methodology allowed for accurate measurement of throughput (ops/sec) and latency variations. Once the YCSB workload executor processed both the data load and run, we analyzed the consistency of the results in MongoDB.

Graphical representations of our findings, detailed in accompanying figures, compare transaction operations across 1, 5, 10, and 15 clients, both pre and post-the-algorithm's integration into the YCSB benchmark. Metrics were captured in time (milliseconds) and operations per second.

We evaluated our proposed algorithm using the YCSB benchmark, focusing on workloads A, B, and F. These benchmarks were chosen to understand how our algorithm impacted throughput, average update latency, and average read latency.

### A. Workload A (Update Heavy Workload)

This workload comprises a 50/50 mix of operations by default. Throughput: As shown in Figure 1, throughput was significantly improved for 1 and 15 clients after applying our algorithm. However, due to possible data inconsistencies during ongoing operations, its effect was comparatively lesser on 5 and 10 clients.

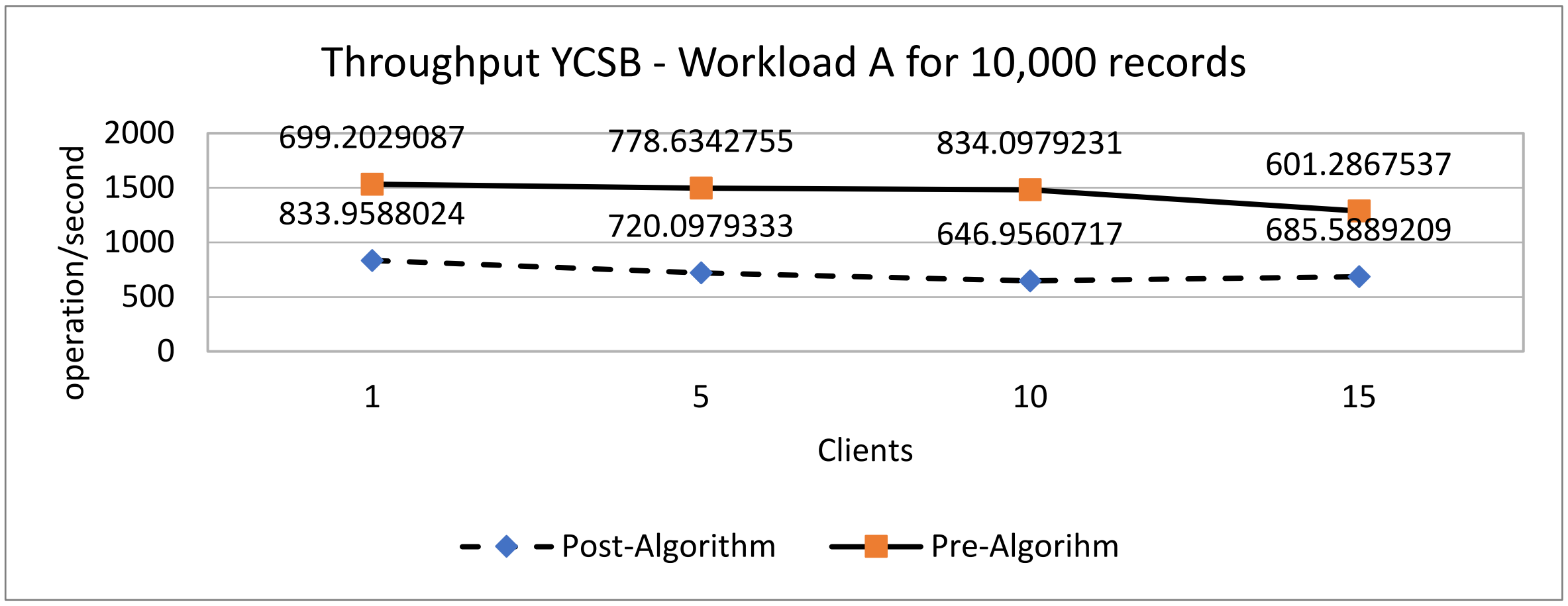

FIGURE 1: COMPARISON OF THROUGHPUT IN YCSB WORKLOAD A FOR 10,000 RECORDS BEFORE AND AFTER IMPLEMENTING THE PROPOSED ALGORITHM.

Update Latency: As presented in Figure 2, data across all clients showed strong consistency after applying the proposed algorithm.

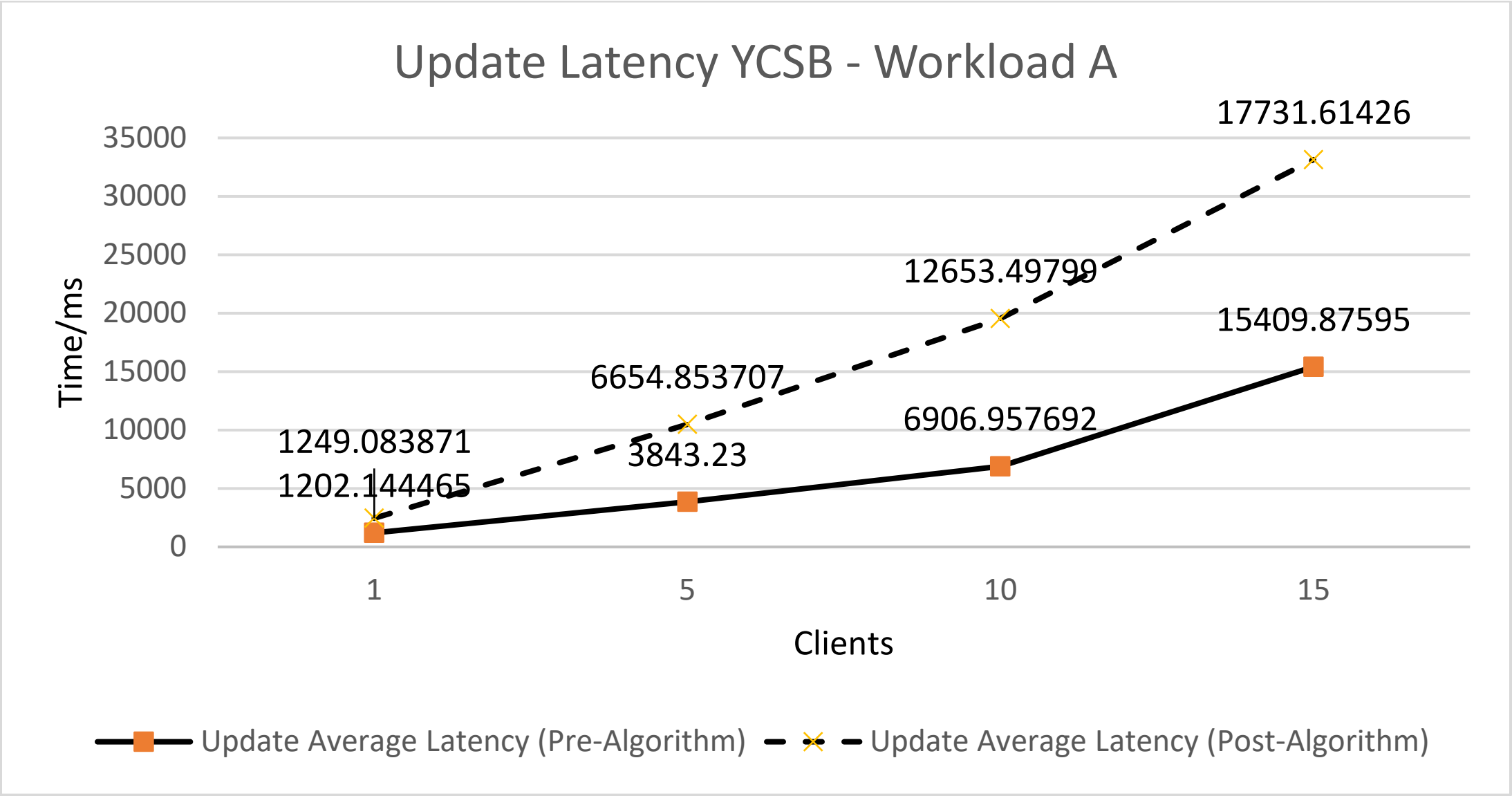

**FIGURE 2: UPDATE LATENCY - WORKLOAD A. 3**

Read Latency: Figure 3 displays the read latency results. Data in 5 and 10 clients showed high consistency, whereas 1 and 15 clients exhibited reduced consistency, which might be attributed to data conflicts when reading multiple operations concurrently.

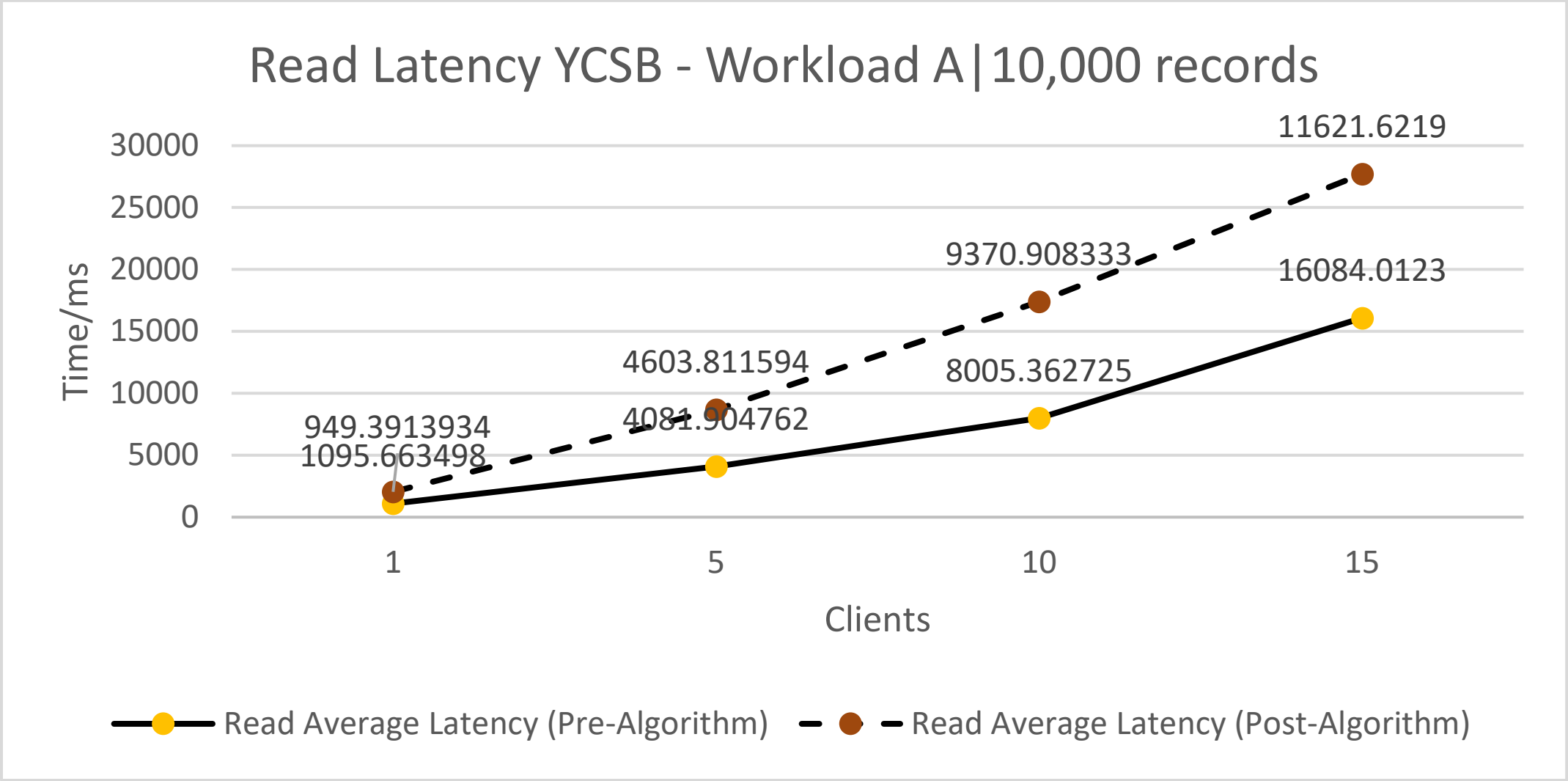

**FIGURE 3:** COMPARISON OF READ LATENCY FOR WORKLOAD A WITH 10,000 RECORDS: PRE-ALGORITHM VS. POST-ALGORITHM IMPLEMENTATION.

Overall, the proposed algorithm enhanced data consistency for Workload A, especially with concurrent transactions. However, the discrepancy in read transactions occasionally led to read failures during execution.

**B. Workload B (Read Mostly Workload):**

This workload consists of a 95/5 read/write mix. Throughput: As evidenced in Figure 4, the throughput was considerably higher for 1, 5, and 15 clients post-algorithm applications. However, the effect was least observed in the 10-client setup due to possible inconsistencies in write operations during execution.

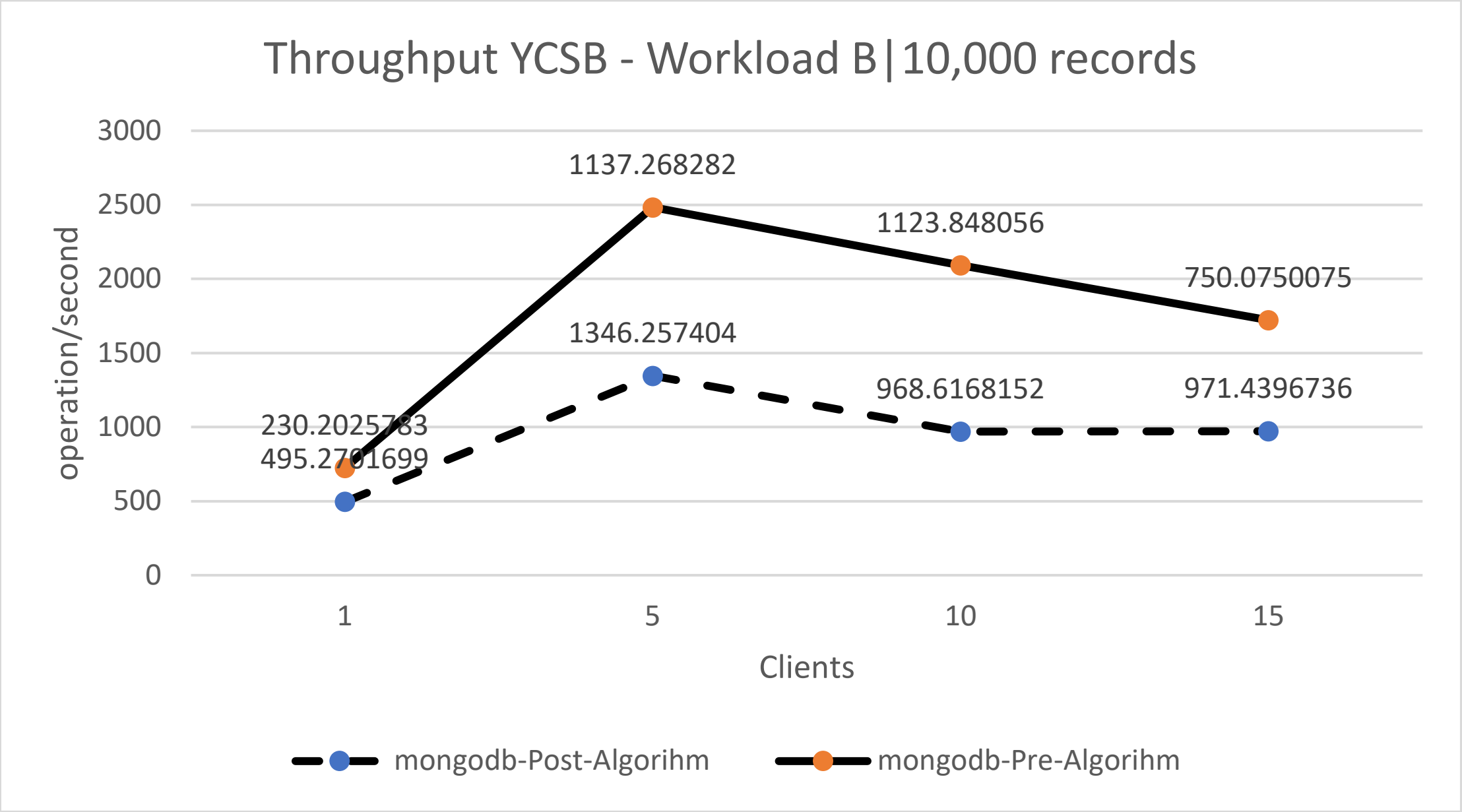

**FIGURE 4:** COMPARISON OF THROUGHPUT IN YCSB WORKLOAD B FOR 10,000 RECORDS BEFORE AND AFTER IMPLEMENTING THE PROPOSED ALGORITHM. THE GRAPH ILLUSTRATES THE NUMBER OF TRANSACTIONS PER SECOND FOR VARIOUS CLIENT CONFIGURATIONS (1, 5, 10, AND 15 CLIENTS) BOTH BEFORE AND AFTER THE APPLICATION OF THE ALGORITHM. THE AIM IS TO HIGHLIGHT THE ALGORITHM'S EFFECTIVENESS IN ENHANCING DATABASE THROUGHPUT.

Update and Read Latency: Both Figure 5 and Figure 6 suggest that our algorithm ensured strong data consistency across the board, with only the 10-client setup showing minor inconsistencies. This pattern is consistent with the algorithm's design, prioritizing strong workloads consistent with predominant read operations.

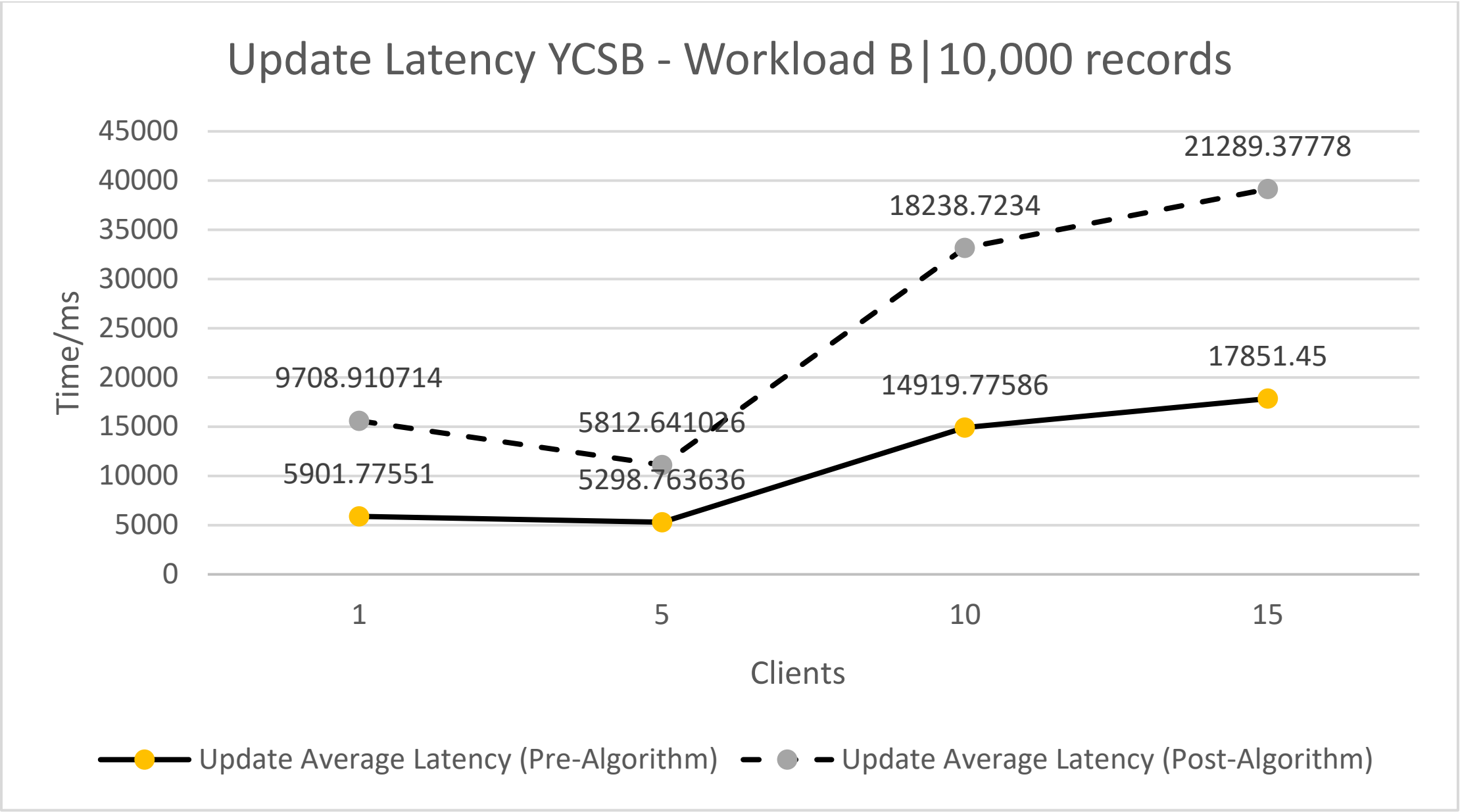

**Figure 5:** Comparison of Update Latency in YCSB Workload B for 10,000 Records Before and After Implementing the Proposed Algorithm

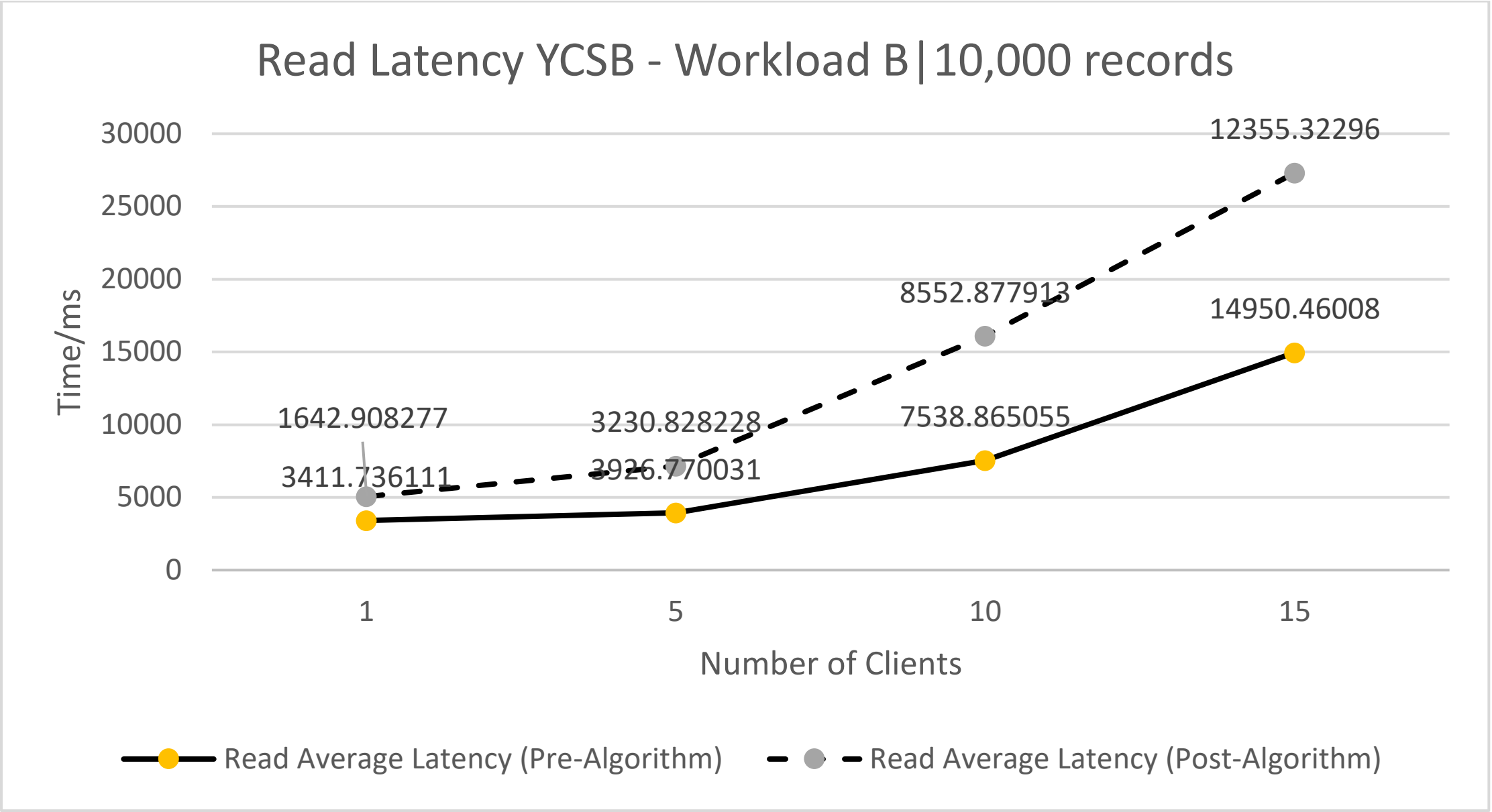

**Figure 6:** Comparison of Read Latency for Workload B with 10,000 Records Before and After Implementing the Proposed Algorithm. This graph illustrates the variation in read latency among 1, 5, 10, and 15 clients, showcasing the impact of the algorithm on improving data consistency and transaction speed.

### C. Workload F (Read, Modify, Write Workload):

Records are read and modified in this workload, and the changes are written back. Throughput: Figure 7 highlights the algorithm's noticeable impact on 10 and 15 clients, with minimal effects on 1 and 5 clients. This could be due to inconsistencies during ongoing operations.

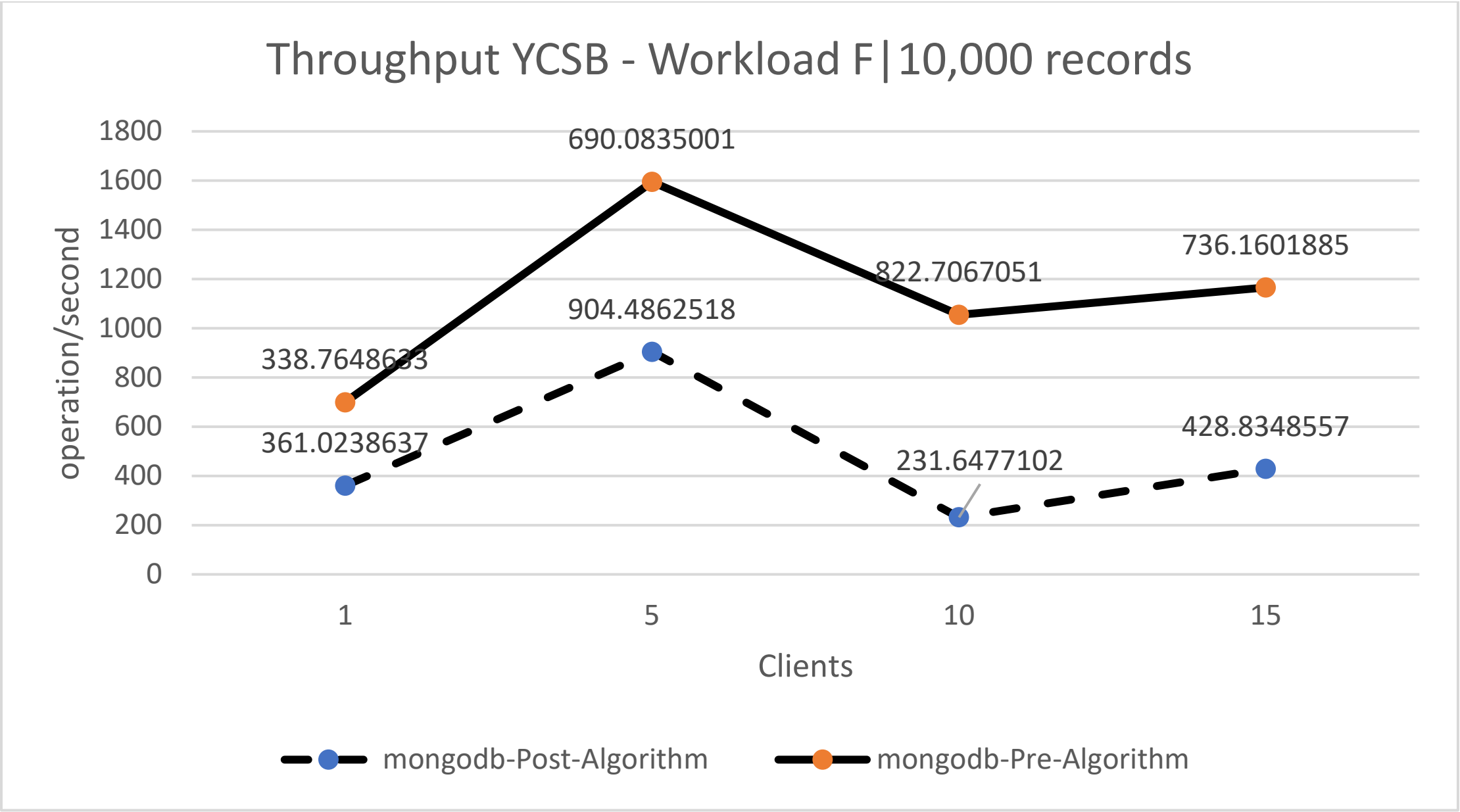

**FIGURE 7:** COMPARISON OF THROUGHPUT IN YCSB WORKLOAD F WITH 10,000 RECORDS, BEFORE AND AFTER IMPLEMENTING THE PROPOSED ALGORITHM. THE GRAPH ILLUSTRATES THE INCREASE IN THROUGHPUT ACROSS DIFFERENT CLIENT SCENARIOS (1, 5, 10, 15 CLIENTS) AFTER THE ALGORITHM'S IMPLEMENTATION, DEMONSTRATING ITS EFFECTIVENESS IN IMPROVING DATABASE PERFORMANCE.

Update Latency: As depicted in Figure 8, most client setups showed increased consistency in update latency post-algorithm application, except for the 5-client setup.

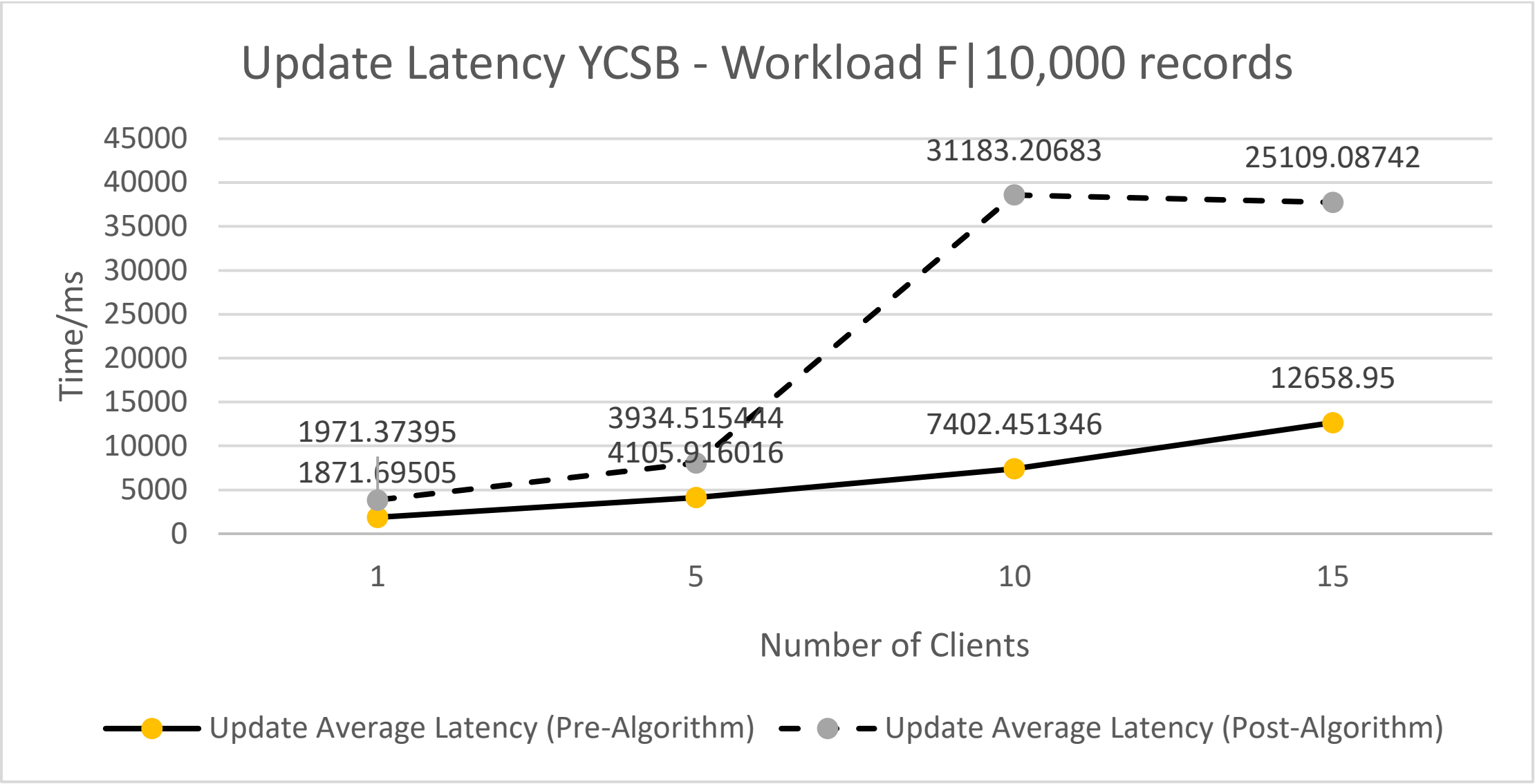

**FIGURE 8:** UPDATE LATENCY YCSB - WORKLOAD F FOR 10,000 RECORDS (PRE-ALGORITHM): THIS GRAPH DEPICTS THE UPDATE LATENCY MEASURED ACROSS VARIOUS CLIENT COUNTS BEFORE THE IMPLEMENTATION OF THE PROPOSED ALGORITHM. THE LATENCY VALUES SERVE AS A BASELINE FOR EVALUATING THE EFFECTIVENESS OF THE NEW ALGORITHM.

Read/Modify/Write Latency: Similar trends were observed in Figure 9 for read/modify/write latency.

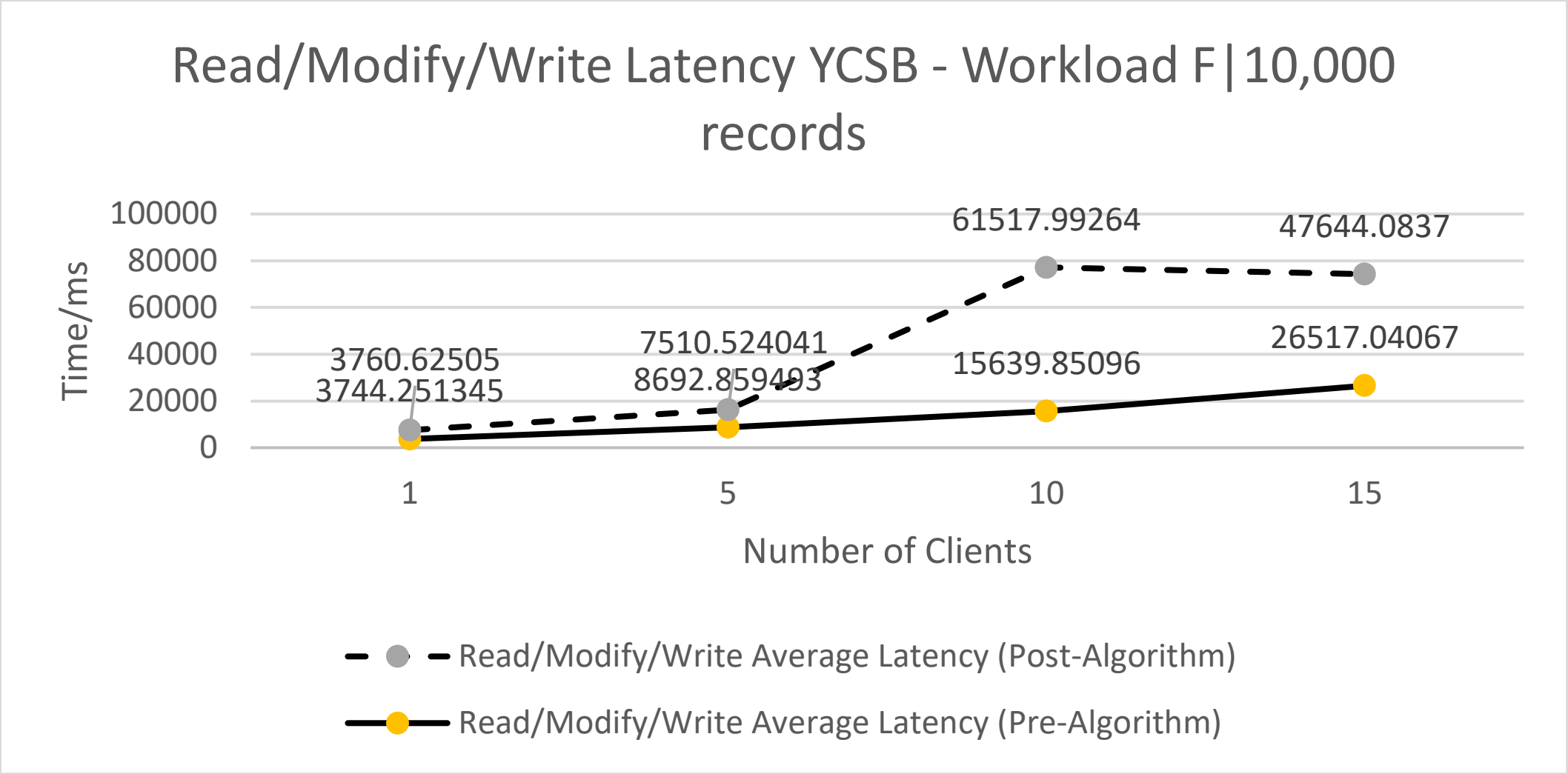

**FIGURE 9:** COMPARISON OF READ/MODIFY/WRITE LATENCY FOR YCSB WORKLOAD F WITH 10,000 RECORDS: PRE-ALGORITHM VS. POST-ALGORITHM IMPLEMENTATION.

Read Latency: In Figure 10, the 10 and 15 client configurations demonstrated the algorithm's impact, with lesser effects on the 1 and 5 client configurations.

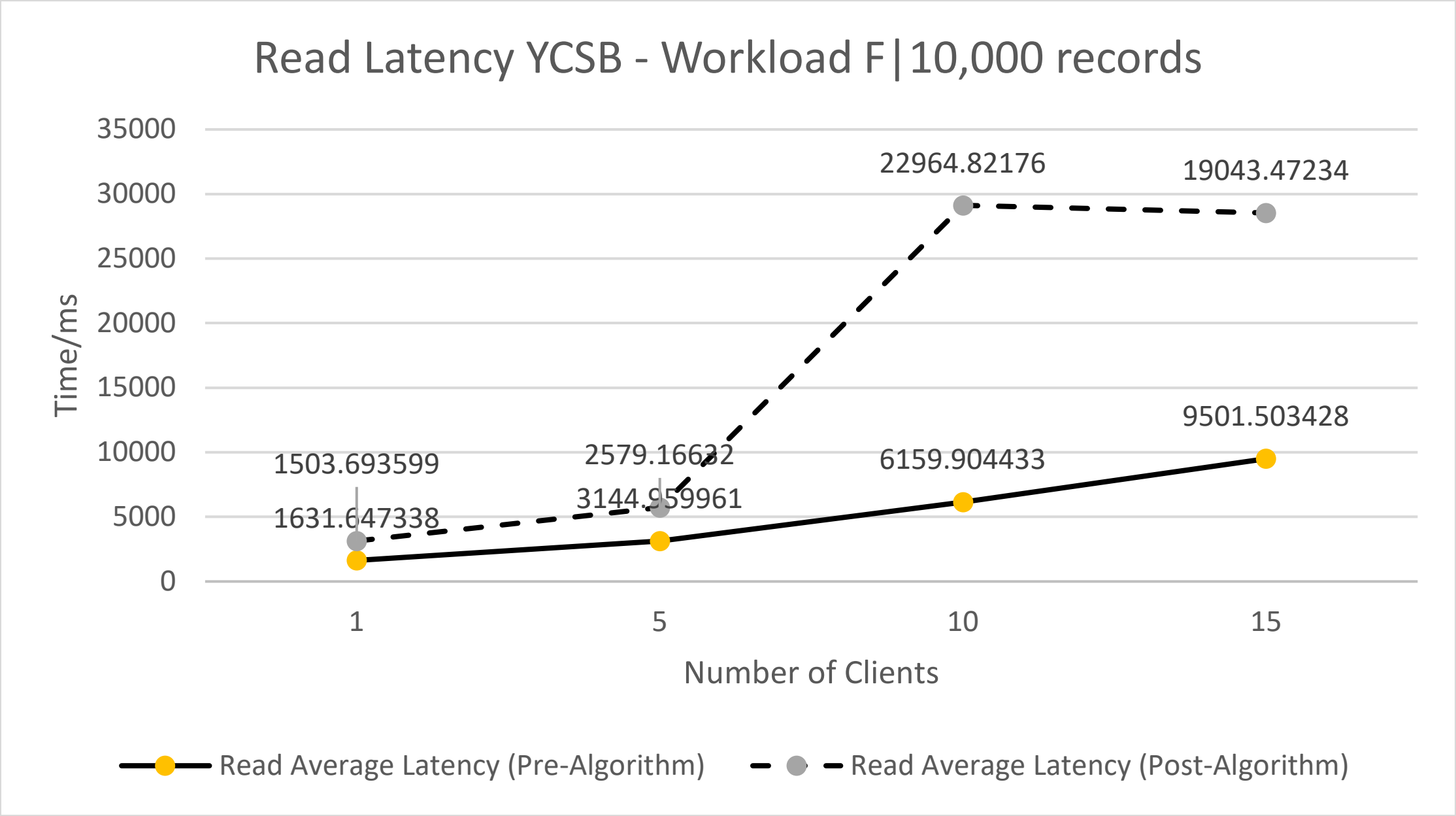

**FIGURE 10:** COMPARISON OF READ LATENCY FOR 10,000 RECORDS IN YCSB WORKLOAD F - PRE-ALGORITHM VS POST-ALGORITHM IMPLEMENTATION.

In summary, for Workload F, our algorithm strengthened consistency in most tests, especially during update operations. However, read operations exhibited varying levels of consistency.

**TABLE 1: AN EFFECT THE PROPOSED ALGORITHM ON MONGODB DATABASE**

| Workloads | Throughput Latency AVG | | | | Update Latency AVG | | | | Read Latency AVG | | | | Read/Modify/Write Latency AVG | | | |
|---|---|---|---|---|---|---|---|---|---|---|---|---|---|---|---|---|
| | Clients | | | | Clients | | | | Clients | | | | Clients | | | |
| | 1 | 5 | 10 | 15 | 1 | 5 | 10 | 15 | 1 | 5 | 10 | 15 | 1 | 5 | 10 | 15 5 |
| **Workload A** | ✓ | ✗ | ✗ | ✓ | ✓ | ✓ | ✓ | ✓ | ✗ | ✓ | ✓ | ✗ | - | - | - | - |
| **Workload B** | ✗ | ✓ | ✗ | ✓ | ✓ | ✓ | ✓ | ✓ | ✗ | ✗ | ✓ | ✗ | - | - | - | - |
| **Workload F** | ✓ | ✓ | ✗ | ✗ | ✓ | ✗ | ✓ | ✓ | ✗ | ✗ | ✓ | ✓ | ✓ | ✗ | ✓ | ✓ |

## 4. Discussion

This section aims to delve deeper into the results garnered using the YCSB benchmark in tandem with the MongoDB database, representing NoSQL databases. Our primary intention was to evaluate the implications of integrating our proposed algorithm within the YCSB benchmark framework, specifically focusing on its potential to enhance data consistency across varying client numbers. The workloads A, B, and F of YCSB provided the foundation for our experimental scenarios, enabling us to gauge average latency throughput, read latency, and update latency.

As illustrated in Table 5.1, there are clear indications of our algorithm's pronounced impact on data consistency within MongoDB. This influence is apparent across a broad spectrum of client data operations, encompassing both read and write functions. One of the standout observations is the distinct enhancement of data consistency, particularly during update operations across most client configurations. Such observations underscore the efficacy of our proposed algorithm in realizing its intended objectives.

However, it's pivotal to recognize that specific client scenarios displayed inconsistencies while the overarching results were positive. This could be attributed to potential inaccuracies in data, mainly when sourced from transactions still in execution. This observation prompts us to exercise caution and underscores the need for further exploration, especially in scenarios where data is being accessed concurrently with its modification.

## 5. Conclusion

In the ever-evolving world of Big Data Processing, NoSQL databases have carved out a significant niche, powering myriad applications ranging from social media platforms like Facebook to expansive online games. Their hallmark is the ability to offer enhanced performance and efficiency tailored to the big data demands inherent in these applications. Yet, a critical challenge these databases face is their inability to guarantee robust consistency during transaction processing.

To address this gap, our research embarked on the ambitious journey of formulating an algorithm designed to bolster data consistency during multi-transaction processes in MongoDB. This devised algorithm underwent rigorous assessment, being integrated within the YCSB benchmark. The intention was to scrutinize its efficacy in ensuring data consistency across various transactional scenarios spanning diverse client configurations, namely 1, 5, 10, and 15 clients. Furthermore, exhaustive tests were executed under YCSB workloads A, B, and F to cater to different client counts.

A comprehensive analysis of the gathered data brought forth promising insights. Incorporating the proposed algorithm led to a noticeable enhancement in data consistency compared to the baseline measurements. Specifically, writing transactions (updates) manifested commendable consistency across most clients. On the contrary, read transactions exhibited a bit of variability in consistency among different clients. However, this variability is deemed acceptable, given that no alterations were made to the data during its processing. In essence, this study underscores the potential of our algorithm as a pivotal tool in mitigating consistency challenges in NoSQL databases.

## 6. Conflict of Interest Statement

The authors declare that there are no conflicts of interest regarding the publication of this paper. All financial, commercial, or other relationships that might be perceived as a potential conflict have been disclosed. The research was conducted without commercial or financial relationships that could be construed as a potential conflict of interest.